\documentclass[%
aip,
amsmath,amssymb,
reprint,%        
]{revtex4-2}

\usepackage{graphicx}%
\usepackage{dcolumn}% 
\usepackage{bm}% 

\usepackage[utf8]{inputenc}
\usepackage[T1]{fontenc}
\usepackage{mathptmx}
\usepackage{etoolbox}
\usepackage{upgreek}

\makeatletter
\def\@email#1#2{%
	\endgroup
	\patchcmd{\titleblock@produce}
	{\frontmatter@RRAPformat}
	{\frontmatter@RRAPformat{\produce@RRAP{*#1\href{mailto:#2}{#2}}}\frontmatter@RRAPformat}
	{}{}
}%
\makeatother

\begin{document}

	\title[Additive solid immersion lenses for enhanced collection efficiency of shallow NV centers]{Additive solid immersion lenses for enhanced collection efficiency of shallow NV centers by pulsed laser deposition and structurization of high-k amorphous oxides.}
	\author{M. S. Bar}
	\affiliation{Felix Bloch Institute, Leipzig University, Linnéstr. 5, 04103 Leipzig, Germany}
	\affiliation{SaxonQ GmbH, Emilienstr. 5, 04107 Leipzig, Germany}%
	\email{michael\_sebastian.bar@uni-leipzig.de}
	\author{N. Raatz}%
	\affiliation{SaxonQ GmbH, Emilienstr. 5, 04107 Leipzig, Germany}%
	\author{M. Kneiss}
	\affiliation{SaxonQ GmbH, Emilienstr. 5, 04107 Leipzig, Germany}%
	\author{A. Koenig}
	\affiliation{Department of Prosthetic Dentistry and Dental Material Science, Leipzig University, Liebigstr. 12, 04103, Leipzig, Germany}%
	\author{M. Grundmann}
	\affiliation{Felix Bloch Institute, Leipzig University, Linnéstr. 5, 04103 Leipzig, Germany}
	\affiliation{SaxonQ GmbH, Emilienstr. 5, 04107 Leipzig, Germany}%
	\author{H. von Wenckstern}
	\affiliation{Felix Bloch Institute, Leipzig University, Linnéstr. 5, 04103 Leipzig, Germany}
	
	\date{\today}
	
	\begin{abstract}
	The negatively charged nitrogen-vacancy ($\mathrm{NV}^-$) defect in diamond can be used for quantum information processing via the readout of the radiative decay of the excited state into the ground state. The large refractive index mismatch of diamond and air causes low photon extraction efficiency due to total internal reflection losses. Here, we investigate the photon extraction numerically and report additive solid immersion lenses (SILs) consisting of the high-k dielectric yttria-stabilized-zirconia (YSZ) for surface modification and enhanced light extraction of shallow implanted nitrogen-vacancy centers. We demonstrate an average increase of 63\% in saturation intensity using a scalable bottum up room temperature pulsed laser deposition and photolithography structurization process. The average necessary laser intensity for saturation is reduced by 33\% by the SILs.
	\end{abstract}
	
	\maketitle

	\section{\label{sec:Intro}Introduction}
	
	Quantum information processing is currently pursued through various qubit modalities and platform technologies, with each platform trying to present higher controllable qubit numbers and fidelity beyond error correction \cite{Chen2024, Bravyi2024, Abobeih2022}. For each approach, unique challenges arise to achieve this goal.   
	The platform technology nuclear spin qubits in diamond, controlled by the electron spin of nitrogen-vacancy centers (NV centers), offers the advantage of being operational under ambient conditions while still achieving leading gate fidelities \cite{Bartling2025}. This reduces requirements for external periphery.
	Initialization and control of the NV center state is typically implemented via laser and microwave pulses while the radiative transition can be used for the readout \cite{Pezzagna2021}. An inherent limitation of this, is the fact that diamond has one of the highest refractive indices in the visible range \cite{Phillip1964, Taylor2023}.  This results in only a small cone (half opening angle of $\phi_c\approx 24.5 ^\circ$) of light originating from a point source that can escape from diamond while the rest is reflected internally.  Various strategies exist for increasing the outcoupling of light and photon extraction efficiency \cite{Barnes2002, Englund2015, Castelletto2011}.  
	For diamond, this is usually achieved by modifying the diamond itself, shaping for example solid immersion lenses (SILs) with focused ion beam (FIB) milling or etching the material into nanobeams or gratings \cite{Marseglia2011, Hadden2010, Lukin2015, Li2015}. This is only feasible if the NV center is far below the surface as defined by the SIL geometry, typically several hundred nanometers if not several micrometers. For high resolution placement of NV centers by ion implantation, the ion energy must be small in the range of 10 keV and subsequently the NV centers are „shallow“ and just 10-20 nm below the surface \cite{Pezzagna2011}. Thus, an additive SIL is called for that is here shaped from an oxide thin film, deposited post implantation on top of the diamond.
	An ex-situ approach has been reported using an top-down approach and individual placement of GaN SILs \cite{Strain2023}. In this work we explore a scalable bottom-up approach using structured pulsed laser deposition of yttria stabilized zirconia (YSZ) as a high-k transparent index matched dielectric on diamond.

	\section{\label{sec:ModellingResults}Influence of refractive index and nonidealities}
	\begin{figure*}[t]
		\includegraphics[width=0.8\textwidth]{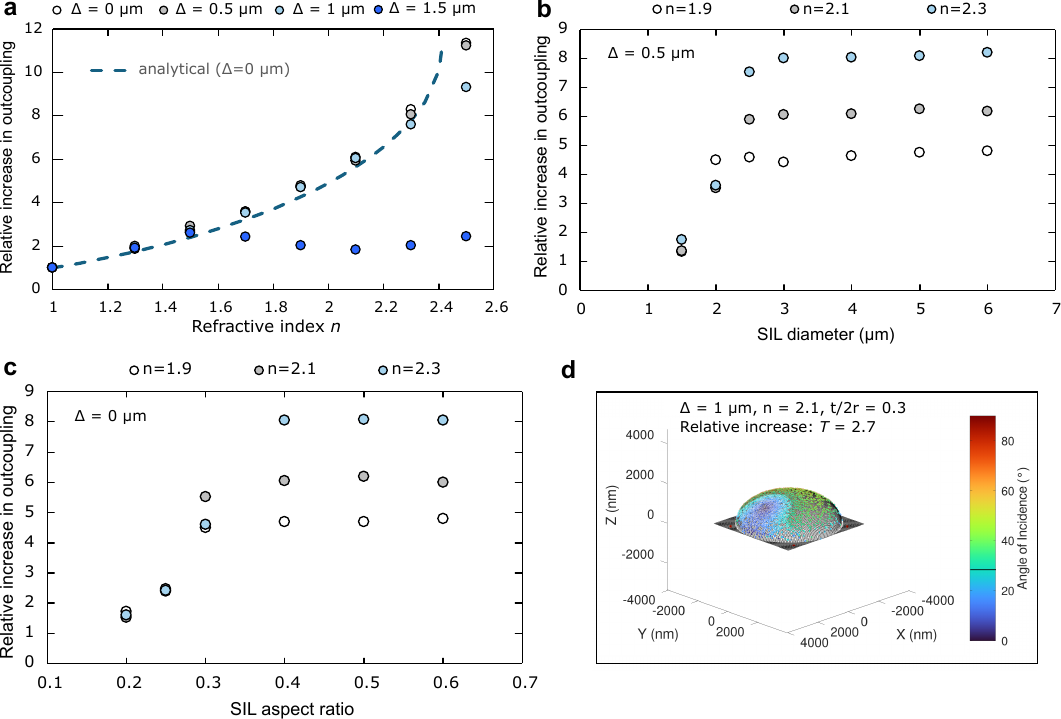}
		\caption{\label{fig:1}a) Increase in light outcoupling from a solid medium as a function of the refractive index of a hemispherical SIL with a diameter of $2r = 5$$\upmu$m. The numerical solutions (circles) are calculated for different lateral offsets $\Delta$ of the point source from the center of the SIL. b) Numerical solution for the light extraction enhancement for a hemispherical SIL with constant offset $\Delta=0.5$$\upmu$m and as a function of total SIL size. c) Numerical solution for the light extraction enhancement as a function of aspect ratio height/diameter. d) An example simulation for a flattened misplaced SIL, still resulting in a twofold increase in outcoupling. The points on the surface show the intersection of rays with the SIL with the color corresponding to the angle of incidence. The critical angle is marked on the colorbar. Outcoupling positions are highlighted with white contour.}
	\end{figure*}
	We performed Monte-Carlo simulations via geometric optics to gain a first understanding of how a certain shape and refractive index improves the outcoupling efficiency from high refractive index materials. The goal of the simulations is to determine how surface geometries affect the total availability of generated photons outside the solid and to systematically introduce nonidealities into our geometry. Initially we considered a SIL of a perfect hemisphere with a refractive index n, a diameter of $2r = 5$$\upmu$m and a thickness of $t = 2.5$$\upmu$m on a 10 nm deep light source. For a perfectly centered homogeneous point source the photon extraction efficiency can then be described as: $\eta(n)=\int_{0}^{2pi}\int_{0}^{\arcsin(n/n_\mathrm{Diamond})}I(\theta,\phi)\,\mathrm{d}\theta\mathrm{d}\phi,$ 
	with $I(\theta,\phi)$ corresponding to the chosen normalized emission distribution. This results in a photon extraction efficiency of $\eta_{\mathrm{analytical}}=4.51 \%$ while our numerical approach for $N = 50 000$ achieves $\eta_{\mathrm{numerical}}=4.52 \%$. Please note that both results do not consider Fresnel reflections in this case and are consistent with previous results \cite{Schubert2006, Barnes2002, Castelletto2011}. In Figure \ref{fig:1}a we show the relative increase in extracted photons $T$ in dependence on the refractive index of the SIL, with both the numerical and analytical solution reaching a tenfold increase for refractive indices of $n>2.4$. For high-k transparent dielectrics with $n \approx 2.1$ we see a sixfold increase in $T$. If the SIL is not placed perfectly centered on top of the point source but with a certain lateral offset $\Delta$ compared to the center of the base the angle of incidence of light with the lens surface is changed. We see that for $\Delta<1$$\upmu$m the original increase is retained. However, for an offset of $\Delta=1.5$$\upmu$m the outcoupling is significantly reduced for $n>1.5$, limiting the increase under $T<3$. Since this is caused by the change in diffraction angles between the light and lens surface this should be considered as a relative error between $\Delta$ and $2r$. This is demonstrated in Figure 1b where the photon extraction efficiency collapses for a fixed offset of $\Delta=0.5$$\upmu$m and a SIL diameter $<2.5$$\upmu$m for $n = 2.3$. For $n = 1.9$ the original increase can be retained into smaller lenses since the total reflection angle is lower. Lastly, we were interested in how a deviation from the ideal hemispherical geometry might affect the light photon extraction efficiency. For this we changed the aspect ratio, where an aspect ratio of $t/2r = 0.5$ corresponds to the ideal shape and lower values represent a flattened lens. The results are depicted in Figure 1c. Similar to before we see that the original gain is retained for aspect-ratio-changes up to $20$ \%. For even smaller aspect ratios we see that T decreases again, with SILs of higher refractive index being more susceptible to changes in geometry like before. A combined mismatch example is shown in Figure 1d.
	Considering these results we can derive the following design considerations. First: High-k dielectrics are highly suitable for SILs even if they are more susceptible to nonidealities since the baseline increase in light collection is highest for refractive index matching. Second: Since standard I-beam lithography systems are limited to a reproducible lateral alignment of 500 nm - 1000 nm one should consider a SIL with a diameter of $2r = 4$$\upmu$m or more. This also implies that the resulting thickness of the deposited dielectric layer should be at least 2$\upmu$m. 
	
	\section{\label{sec:Deposition}Layer Deposition and Structurization}
	Material choice is an important factor for the development of additive SILs. In the case of optical readout of NV centers the prerequisites are sufficient growth rate and adhesion on diamond, high refractive index and transparency in the visible range and low autofluorescence for the NV center excitation wavelength. One material that fulfills these conditions is room temperature pulsed laser deposited yttria stabilized zirconia (YSZ), using targets of partially stabilized tetragonal zirconia (3Y-TZP) as source material \cite{Lippert2010, Lippert2011}.
	A deposition of 500 000 Pulses at a 20 Hz repetition rate and a fluence of $2.75\,\mathrm{J/cm}^2$ at room temperature and an oxygen background pressure of 0.01 mbar results in a thin film of approximately 2.5$\upmu$m thickness. The refractive index of the layer for 632 nm was determined to be $n=2.17$ with a Cauchy model, which is applicable for up to 4.3 eV, which is where the amorphous absorption onset begins. In Figure 2 we show the fluorescence as a function of laser power which shows comparable intensity for a nominally undoped diamond and the YSZ layer.
	A previously proposed top-down approach is not suitable for the structurization of YSZ due to the high etch resistance of the material \cite{Sardi2020, Park2026}. We therefore demonstrate a simple scalable bottom-up proof of concept for the cost-effective fabrication of angled deposition solid immersion lenses (AD-SILs). We developed holes with a nominal size of $2r=6$$\upmu$m into a 5.5$\upmu$m photoresist layer, resulting in an aspect ratio close to one. Subsequent structurization is realized during deposition by an offset between target and substrate (see Figure 3a). Due to this, only a part of the substrate is exposed to the material flux. Rotating the sample during the process results in a curvature of the surface.
	The resulting array can be seen in Figure 3b with a close-up cross section of a single SIL in Figure 3c. Clearly visible in the array as well as the close-up is the individual curvature as well as a surrounding “eggshell” like ring. This is a result of deposition of material onto the photoresist sidewalls which is then not irregularly removed during lift-off, also known as “fencing”. Further lithographic processing optimization might mitigate this behavior. It should be noted that slight fencing might however still occur if the angle of deposition exceeds the photoresist sidewall retrograde. It is also not clear whether this affects the performance of the SIL itself.
	\begin{figure}[t]
		\includegraphics[width=0.9\columnwidth]{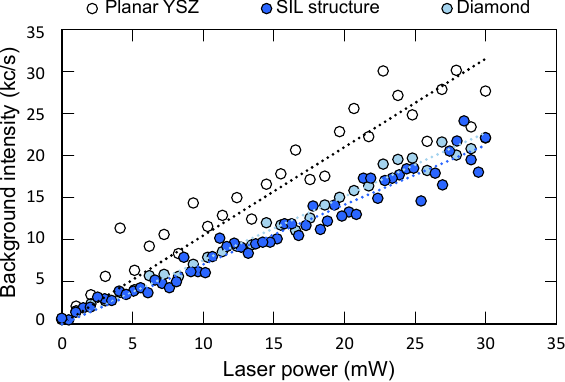}
		\caption{\label{fig:epsart} Luminesence of the deposited YSZ layer as a function of excitation power with an electronic grade diamond as reference. The dashed line represents a linear fit.}
	\end{figure}
	
	\begin{figure*}[t]
		\includegraphics[width=0.7\textwidth]{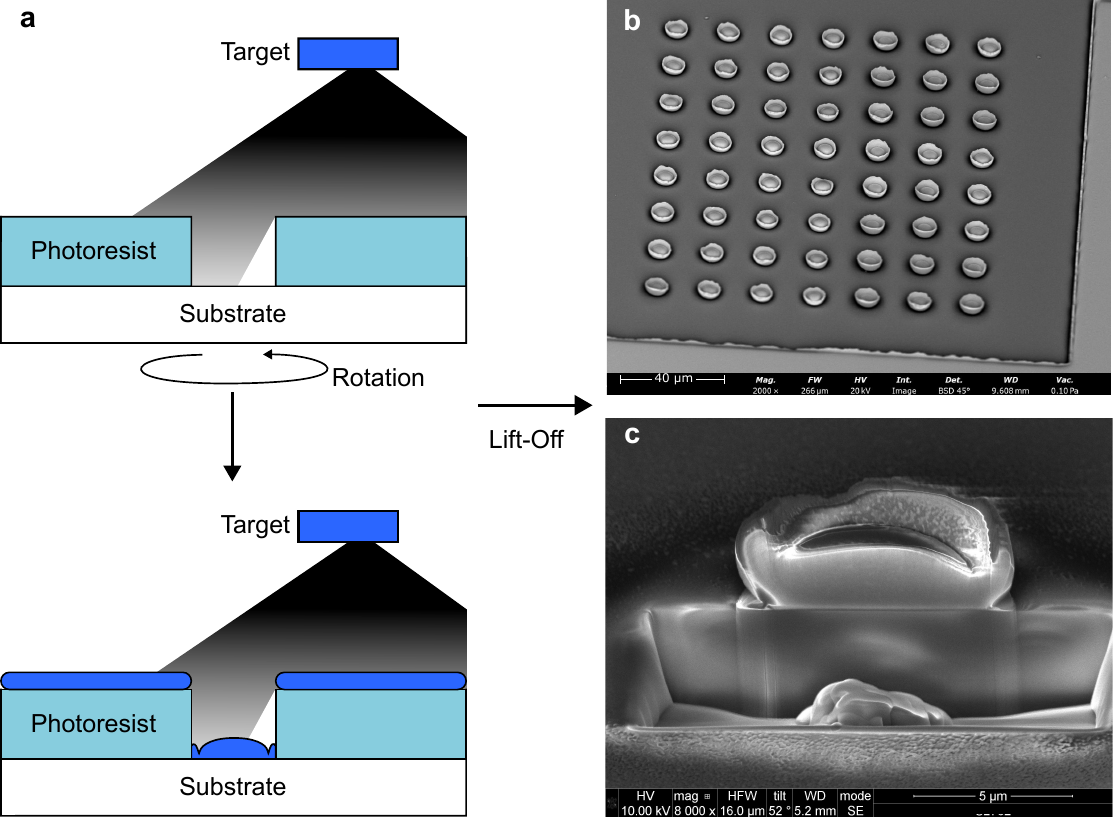}
		\caption{\label{fig:3}a) Schematic illustration of the deposition process of AD-SILs. b) Array of YSZ AD-SIL fabricated simultaneously with PLD on glass. c) Cross-section of a YSZ AD-SIL on glass with “fencing”.}
	\end{figure*}
	\section{\label{sec:NVstats}NV center statistics}
		\begin{figure*}[t]
		\includegraphics[width=0.8\textwidth]{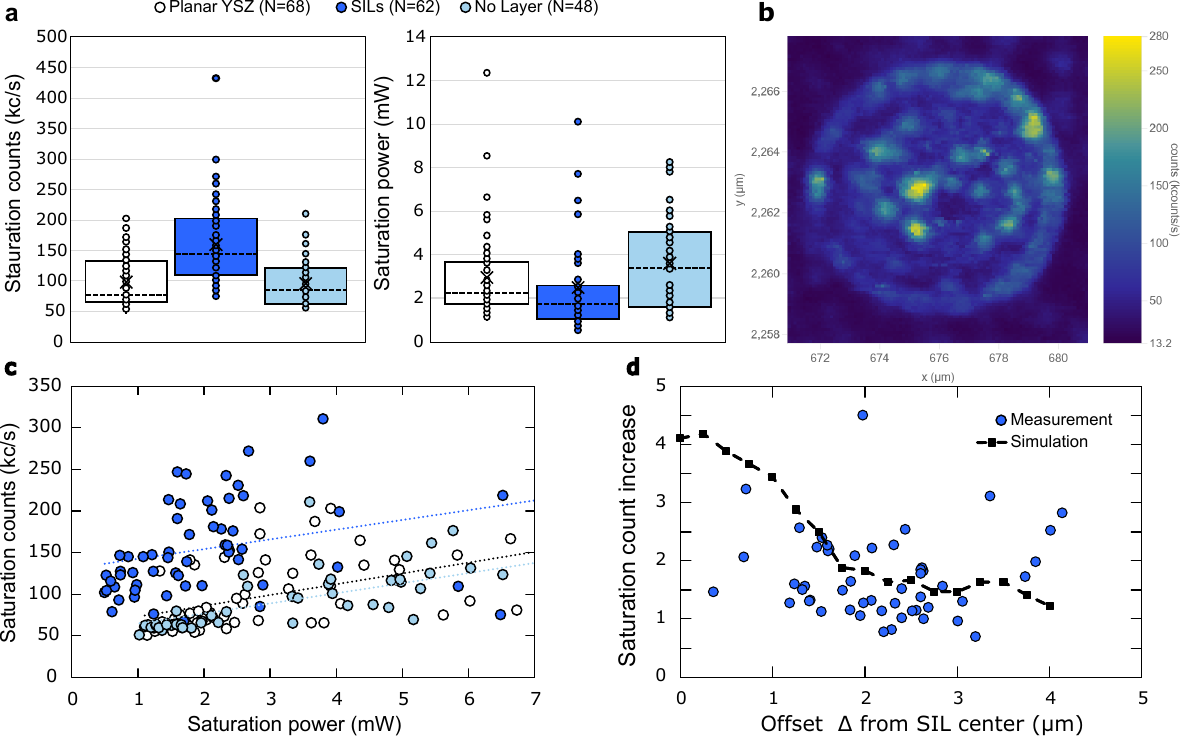}% Here is how to import EPS art
		\caption{\label{fig:4}a) Statistical evaluation of the saturation intensity and saturation power of NV centers under planar YSZ, AD-SILs and bare diamond surface. The circles denote individual measurements, the box the first and third quartile and the dashed line and cross the median and average respectively. b) Example confocal scan of an AD-SIL on diamond with visible NV center luminescence. c) Saturation counts as a function of respective power for the respective surface modification types. The dashed lines are linear fits meant as a guide to the eye. d) Measured and simulated saturation count increase of NV centers under AD-SILs compared to the average uncovered NV center. }
	\end{figure*}
	To determine the effectiveness of the SIL we evaluated the background corrected saturation counts and power of individual $\mathrm{NV}^-$ centers under SILs, under planar YSZ and untreated diamond. This rules out effects due to potential surface passivation \cite{Neethirajan2023}. We determined the saturation according to $I(P)=I_\mathrm{sat}/(1+P_\mathrm{sat}/P)$ with $P$ being the nominal laser power \cite{Neu2011}.
	It should be noted that the evaluation presented here contains measurements with and without a sulfur background doping, which is known to affect the charge state of the NV center \cite{Luhmann2019, Becker2023}. Furthermore, measurements of single NV centers as well as pairs are included. We therefore expect to observe a spread in luminescence and saturation intensity. We were however interested in the resulting statistical effect of our SILs which can be seen in Figure 4a with an example overview scan of NV centers under a SIL in Figure 4b. 
	For NV centers under planar YSZ and bare diamond we observe an average saturation count of $\overline{I_{\mathrm{sat}}}=98\,\mathrm{kc/s}$ and $\overline{I_{\mathrm{sat}}}=96\,\mathrm{kc/s}$ respectively. Compared to the latter, the luminescence of NV centers located under SILs increases on average by 63\% to $\overline{I_{\mathrm{sat}}}=160\,\mathrm{kc/s}$. A similar increase is observed when considering the median. The aforementioned spread in saturation intensity is increased as the measured luminescence now additionally depends on the SIL effectiveness.  Simultaneously we observe a 33\% decrease in average saturation power from $\overline{P_{\mathrm{sat}}}=3.6\,\mathrm{mW}$ (no deposition) and $\overline{P_{\mathrm{sat}}}=3.0\,\mathrm{mW}$ (planar YSZ)  down to $\overline{P_{\mathrm{sat}}}=2.5\,\mathrm{mW}$. These effects are particularly noticeable if one depicts $I_{\mathrm{sat}}$  as a function of $P_{\mathrm{sat}}$ in Figure 4c. 
	In the range of $0 \mathrm{mW}<P_{\mathrm{sat}}<3 \mathrm{mW}$ it is especially prominent that NV centers under SILs reach higher saturation counts for the same or lower excitation power. Specifically, NV centers under SILs reach $I_{\mathrm{sat}}$ values above 200 kc/s and $P_{\mathrm{sat}}$ values below 1 mW which is not observable for either of the other cases. In contrast to the intensity the change in saturation power is not fully intuitive. One may assume that the power density of excitation increases due to the SIL, but this is only reasonable if the point spread function degradation due to high refractive index of diamond is somewhat mitigated due to the non-planar incident angle \cite{EgnerHell1999}. The incident angle change might however also affect the polarization of the excitation beam compared to the planar case which is why this phenomenon is subject to further investigations.
	We measured the surface morphology of a comparable AD-SIL using atomic force microscopy and used the data for the simulation discussed in Section II in order to estimate the previously discussed misplacement effect. At the same time, we compare the results with the measured results in Figure 4d. From the simulation we see the highest increase in outcoupling for centrally placed NV centers with a noticeable dropoff for an increase of $\Delta$ from 1 to 2$\upmu$m . For  $\Delta>2\,\upmu $m  the dropoff continues, albeit less severe. It is difficult to compare the measured case for centrally placed NV centers with the simulation as the number of centrally placed NVs is considerable less statistically relevant in the experiment. Most NV centers were found between 1 and 3 µm from the center, for which a good  agreement between model and experiment was observed.. For NV centers further from the center edge effects must be taken into accounts due to the fencing of the SIL.
	\section{\label{sec:Conclusion}Conclusion and Outlook}
	We demonstrate a simple approach to light collection challenge from diamond NV centers. Geometric modeling shows that high-k dielectrics provide measurable collection gains despite imperfect SIL geometries. A larger SIL is more robust towards misplacements. Our findings show that 3Y-TZP targets and the resulting films deposited by PLD are highly suitable for surface modification diamonds due to the high transparency, high refractive index and low autofluorescence. We further demonstrate this by employing a simple bottom-up deposition of a SIL array on shallow NV centers, resulting in an average increase in saturation intensity of 63 \%.
	We believe that this gain can be further increased with improved hemispherical geometries either by an improved fine tuning the angle between substrate and target or by individual FIB milling of selected positions. The modelling and effect of the geometry would benefit from a numerical wave optics approach that further also considers the non-uniformity of the NV center emission distribution \cite{Dolan2014}. The deposition of YSZ should also be fully compatible with the implementation of a $\mathrm{TiO_2}$ interlayer which may improve charge stability and NV center fluorescence even further \cite{Yu2026}.

	\section{\label{sec:Experimental}Experimental}
	The pulsed laser deposition was performed with a \textit{Lambda Physik LPX Pro} KrF 248nm excimer laser. Details of the PLD Setup can be found in \cite{Lorenz2015}. The YSZ target was fabricated from tetragonal partially stabilized zirconia (3Y-TZP) made by \textit{Dental Direkt}. Lithography was performed with a \textit{Heidelberg Instruments $\mu$MLA} maskless aligner. Layer thickness was resolved using a \textit{Bruker Dektakt XT} stylus profilometer. The refractive index was determined using a \textit{J. A. Woollam RC2} ellipsometer. Scanning electron microscopy was performed with \textit{Thermo Fischer Phenom XL} and \textit{FEI Nova Nanolab 200}. The latter was also used for focused ion beam milling. Atomic force microscopy measurements were performed using \textit{Park Systems AFM XE-150}. The measurement of the luminescence was carried out with a custom-built confocal scanning setup using a 514 nm excitation wavelength and a 0.95 NA objective.\\
	The electronic grade (100)-oriented diamond from \textit{ElementSix} was first co-doped with sulfur by using 10 keV $\mathrm{S}^-$ ions at the \textit{Leipzig University 100 keV accelerator} and a following annealing step of 1200$^\circ$C for 4h in a high vacuum oven. Details of the effect of co-doping can be found in\cite{Luhmann2019, Becker2023}. The NV Centers were then created by ion implantation with 10keV $\mathrm{N_2}^{2+}$ ions in the \textit{Leipzig University Nanoimplanter}\cite{Raatz2019} setup and activated through a second annealing procedure at 800$^\circ$C for 2h. In a final step the diamond was cleaned with a soft oxygen plasma in a \textit{Diener Pico} low pressure plasma chamber for 10min at 200$^\circ$C.
	
	\section{\label{sec:Modeling}Modeling}
	For diamond we considered $n_\mathrm{Diamond}=2.41$ \cite{Phillip1964, Taylor2023}.  This simulation tracks homogeneous generated photons from a point source. The light is considered unpolarized. A photon is considered collected if it passes the interface between lens and air. At every interface we consider diffraction, total- and Fresnel reflection. The intersection of the photons with an arbitrary lens surface is calculated numerically. A photon is considered lost if it is totally reflected at any of the interfaces or misses the SIL. We do not consider multiple internal reflections, wave optics or the numerical aperture of our optics setup. The maximum efficiency achievable with these presumptions can therefore be 50\% of all generated photons. The simulation was written in \textit{MATLAB}/\textit{GNU Octave}.
	\begin{acknowledgments}
		This work was funded by the DATIpilot project \textit{QpDiR-Q} (grant nr: 03DPS1123A) granted by the German Federal Ministry of Research, Technology and Space. We thank Fabian Schöppach for the help with the preparation of PLD targets and Daniel Splith  for the help with the modelling (both Felix Bloch Institute, Leipzig University).
	\end{acknowledgments}

		\bibliography{Additive-SILs}%

	\end{document}